

\input harvmac.tex

\noblackbox
\def\ie{{\it i.e.}}

\def\half{{1 \over 2}}

\def\punct[#1]{{\hbox{\hskip .5cm #1}}}

\def\np#1#2#3{Nucl. Phys. {\bf #1} (#2) #3}
\def\pl#1#2#3{Phys. Lett. {\bf #1} (#2) #3}
\def\prl#1#2#3{Phys. Rev. Lett. {\bf #1} (#2) #3}

\def\prd#1#2#3{Phys. Rev. D {\bf #1} (#2) #3}

\def\Boxmark#1#2#3{\global\setbox0=\hbox{\lower#1em \vbox{\hrule height#2em
     \hbox{\vrule width#2em height#3em \kern#3em \vrule width#2em}%
     \hrule height#2em}}%
     \dimen0=#2em \advance\dimen0 by#2em \advance\dimen0 by#3em
     \wd0=\dimen0 \ht0=\dimen0 \dp0=0pt
     \mkern1.5mu \box0 \mkern1.5mu }

\Title{\vbox{\baselineskip12pt\hbox{BUHEP-92-23}\hbox{hep-ph/9206255}}}
{\vbox{\centerline{The Phenomenology of a Hidden}
	\vskip2pt\centerline{Symmetry Breaking Sector}}}

\centerline{R. Sekhar Chivukula, Mitchell Golden,
Dimitrios Kominis, and M. V. Ramana$^*$}
\vskip .1in
\centerline{\it Boston University}
\centerline{\it Dept. of Physics}
\centerline{\it 590 Commonwealth Avenue}
\centerline{\it Boston, MA 02215}
\vskip .2in
\centerline{\bf ABSTRACT}

We calculate the production rate of gauge-boson pairs at the SSC in a model
with a ``hidden'' electroweak symmetry breaking sector. We show that the
signal of electroweak symmetry breaking is lower than the background
and that we cannot necessarily rely on gauge boson pairs as a signal of
the dynamics of symmetry breaking.

\footnote{}{$^*$ sekhar@weyl.bu.edu, golden@weyl.bu.edu,
kominis@budoe.bu.edu, mani@ryan.bu.edu}

\Date{06/92}

It is generally assumed that in the elastic scattering of longitudinally
polarized $W$s and $Z$s either there will be resonances with masses of less
than 1 TeV (as in the weakly coupled one doublet Higgs model) or the
scattering amplitudes will become large, indicating the presence of new strong
interactions at or above 1 TeV (see, for example \ref\Chan{M.~S. Chanowitz,
Ann.  Rev. Nucl. Part. Sci. {\bf 38} (1988) 323 \semi M.~Golden, in {\it
Beyond the Standard Model, Iowa State University, Nov. 18-20, 1988}, K.
Whisnant and B.-L. Young Eds., p. 111, World Scientific, Singapore, 1989.}).
In a recent paper \ref\hidden{R.~S.~Chivukula and M.~Golden, Phys.  Lett. {\bf
B267} (1991) 233.}, two of us argued that there is another possibility: If the
electroweak symmetry breaking sector has a large numbers of particles other
than the longitudinal components of the $W$ and $Z$, then the {\it elastic}
$W$ and $Z$ scattering amplitudes can be small and structureless, i.e. lacking
any discernible resonances, at all energies.  It was further argued that in
such a model it may not be possible to rely on the two-gauge-boson events as a
signal of the symmetry breaking sector. A toy model with these properties,
based on an $O(N)$ scalar field theory solved in the limit of large
$N$, was discussed.

These arguments have been recently been disputed by Kane, Naculich, and
Yuan \ref\kny{G.~L.~Kane, S.~G.~Naculich, and C.~P.~Yuan, ``Effects of
Inelastic Channels on Experimental Detection of the Electroweak Symmetry
Breaking Sector'', University of Michigan preprint UM-TH-92-03\semi
S.~G.~Naculich and C.~P.~Yuan, ``Can the Electroweak Symmetry-breaking
Sector Be Hidden?'', Johns Hopkins University Preprint JHU-TIPAC-920017.}
who argue that, {\it independent of N}, the {\it total} number of $W_L
W_L,\ Z_L Z_L \to Z_L Z_L$ scattering events in the $O(N)$ model is
comparable to the number of events due to a standard-model ``TeV Higgs''
boson, and may therefore be large enough to be observed. However, for large
$N$ the would-be Higgs resonance is \hidden\ both light and very broad. In
this note, we compute both the gauge boson pair signal and background for
the toy model presented in ref. \hidden. We show that, while the number of
gauge boson scattering events is approximately independent of $N$, the
background is {\it much} larger for a light resonance and this signal is
not observable.

We also estimate the $Z_L Z_L$ signal which arises from gluon fusion through
a top quark loop \ref\gfusion{H.~Georgi, S.~L.~Glashow, M.~E.~Machacek, and
D.~V.~Nanopoulos, \prl {40} {1978} {692}.}. Because the would-be Higgs of
this model is light and broad, even this contribution to the signal is
smaller than the background by a factor of four or more making detection
problematic, at best.

We begin by reviewing the toy model of the electroweak symmetry sector
constructed in \hidden. This model has both exact Goldstone bosons (which
will represent the longitudinal components of the $W$ and $Z$
\ref\equiv{ M.~S.~Chanowitz and M.~K.~Gaillard, \np {B261} {1985} {379}.})
and pseudo-Goldstone bosons. The Lagrangian density is
\eqn\lnought{
{\cal L} = \half (\partial \vec{\phi})^2 +
\half (\partial \vec{\psi})^2 - \half \mu_{0\phi}^2 \vec{\phi}^2 - \half
\mu_{0\psi}^2 \vec{\psi}^2 - {\lambda_0 \over 8 N} {(\vec{\phi}^2 +
\vec{\psi}^2)}^2
\punct[,]
}
where $\vec{\phi}$ and $\vec{\psi}$ are $j$- and $n$-component real vector
fields. This theory has an approximate $O(j+n)$ symmetry (i.e. $N=j+n$)
which is softly broken to $O(j) \times O(n)$ so long as $\mu_{0\phi}^2
\neq \mu_{0\psi}^2$. If $\mu_{0\phi}^2$ is negative and less than
$\mu_{0\psi}^2$, one of the components of $\vec{\phi}$ gets a vacuum
expectation value (VEV), breaking the approximate $O(N)$ symmetry to
$O(N-1)$. With this choice of parameters, the exact $O(j)$ symmetry is
broken to $O(j-1)$ and the theory has $j-1$ massless Goldstone bosons and
one massive Higgs boson. The $O(n)$ symmetry is unbroken, and there are $n$
degenerate pseudo-Goldstone bosons of mass $m_\psi$ ($m^2_\psi =
\mu^2_{0\psi} - \mu^2_{0\phi}$). This model is particularly interesting
since it can be solved (even for strong coupling) in the limit of large $N$
\ref\Coleman{S.~Coleman, R.~Jackiw, and H.~D.~Politzer, Phys. Rev. {\bf
D10} (1974) 2491.}. We will consider this model in the limit that $j,n
\rightarrow \infty$ with $j/n$ held fixed.

The scalar sector of the standard one-doublet Higgs model has a global $O(4)
\approx SU(2) \times SU(2)$ symmetry, where the 4 of $O(4)$ transforms as
one complex scalar doublet of the $SU(2)_W\times U(1)_Y$ electroweak gauge
interactions. It is this symmetry which is enlarged in the $O(N)$ model: we
will model the scattering amplitudes of longitudinal gauge bosons by the
corresponding $O(j)$ Goldstone boson scattering amplitudes in the $O(j+n)$
model solved in the large $j$ and $n$ limit. Of course, $j=4$ is not
particularly large. Nonetheless, the resulting model will have all of the
correct qualitative features, the Goldstone boson scattering amplitudes will
be unitary (to the appropriate order in $1/j$ and $1/n$), and we can
investigate the theory at moderate to strong coupling \ref\Einhorn{
M.~B.~Einhorn, Nucl. Phys. {\bf B246} (1984) 75\semi R.~Casalbuoni,
D.~Dominici, and R.~Gatto, Phys. Lett. {\bf 147B} (1984) 419.}. We make no
assumptions about the embedding of $SU(2)_W \times U(1)_Y$ in $O(n)$, \ie\ no
assumptions about the electroweak quantum numbers of the pseudo-Goldstone
bosons: we will assume, however, that the pseudo-Goldstone bosons are $SU(3)$
color singlets\foot{Gauge boson pair production in models with colored
pseudo-Goldstone bosons is discussed in detail in \ref\gscat{J.~Bagger,
S.~Dawson, and G.~Valencia, \prl{67}{1991}{2256}\semi R~S.~Chivukula,
M.~Golden, M.~V.~Ramana, \prl {68}{1992} {2883}.}.}.

One may compute Goldstone boson scattering to leading order in $1/N$. The
details of the calculation may be found in \hidden.  The amplitude
$a^{ij;kl}(s,t,u)$ for the process $\phi^i \phi^j \to \phi^k \phi^l$ is
\eqn\scatamp{
a^{ij;kl}(s,t,u) = A(s)\delta^{ij}\delta^{kl} +
A(t)\delta^{ik}\delta^{jl} +
A(u)\delta^{il}\delta^{jk}
}
where
\eqn\scatII{
A(s) = {s\over
v^2 - Ns\left({1\over\lambda(M)}
+\widetilde{B}(s;m_\psi,M)\right)}
\punct[,]
}
and
\def\xx{\sqrt{s / (4 m_\psi^2 - s)}}
\eqn\Btilde{
\eqalign{
\widetilde{B}(s; m_\psi, M)  = {n \over 32 N \pi^2}
& \left\{
 1
+ {i \over \xx} \log{i - \xx \over i + \xx}
- \log{m_\psi^2 \over M^2}
\right\}\cr
&+
{ j \over 32 N \pi^2 }
\left\{ 1 + \log{M^2 \over -s}
\right\}\punct[.]\cr
}
}
Here $s$, $t$, and $u$ are the usual Mandelstam variables, $v$ is the weak
scale (approximately 250 GeV), and $M$ is a renormalization point which we
chose below such that the renormalized coupling, $\lambda(M)$, satisfies
$1/\lambda(M)=0$.

\nfig\nxxxii{Differential production cross section for $pp \to ZZ$ (at a
$pp$ center of mass energy of 40 TeV) as a function of invariant $Z$-pair
mass for $j=4$, $n=32$, $m_\psi = 125$ GeV and the renormalization point
$M=1500$ GeV. A rapidity cut of $|y| < 2.5$ has been imposed on the final
state $Z$s. The gauge boson scattering signal is shown as the dot-dash
curve and gluon fusion signal (with $m_t = 120$ GeV) as the solid curve.
The background from $q \bar{q}$ annihilation is shown as the dashed curve.
In all contributions, the rapidities of the $Z$s must satisfy $|y_Z|<2.5$.
All computations use the EHLQ set II \ehlq\ structure functions with $Q^2 =
M_W^2$ in the gauge boson scattering curve, and $Q^2 = \hat{s}$ in the
other two cases.}

\nfig\nviii{Same as fig. 1 with $n=8$ and $M=2500$ GeV, as in ref. \kny.}

\nfig\nz{Same as fig. 1 with $n=0$ (ref. \Einhorn) and $M=4300$ GeV.}

\nfig\higgs{Same as fig. 1 for a standard model Higgs boson with mass 485 GeV.}

The amplitude $a^{ij;kl}$ of eqn. \scatamp\ may be used to derive partonic
cross sections for $W_L W_L,\ Z_LZ_L \to Z_LZ_L$ which can then be folded with
the appropriate gauge boson structure functions (using the ``effective-W
approximation'' \ref\effectw{M.~S.~Chanowitz and M.~K.~Gaillard, \pl
{142B}{1984}{85} \semi S.~Dawson, \np {B249}{1985}{42}\semi G.~L.Kane,
W.~W.~Repko, and W.~B.~Rolnick, \prd {10}{1984}{1145}.} and the EHLQ set II
\ref\ehlq{E.~Eichten, I.~Hinchliffe, K.~Lane, and C.~Quigg, Rev. Mod. Phys.
{\bf 56} (1984) 579.} structure functions) to yield the contribution of gauge
boson scattering to the process $pp \to ZZ + X$. This contribution to the
differential cross section for $ZZ$ production as a function of $ZZ$ invariant
mass is shown in the dot-dash lines of \nxxxii\ for $n=32$ and $M=1500$ GeV
\hidden.

As in the standard model, gluon fusion through a top quark loop \gfusion\
provides a signal for gauge boson pairs comparable to the signal from gauge
boson scattering.  The correct computation in this model is somewhat
nontrivial.  We wish to work to lowest nonvanishing order in $\alpha_s$ (the
QCD coupling constant) and the top quark Yukawa coupling.  To this order,
there are three diagrams that contribute to the $Z_LZ_L$ final state.  The
first is the simple top quark triangle diagram, the analogue of the Higgs
production diagram in the standard model.  Next there is the top-quark box,
which produces final state longitudinal $Z$'s exactly as in the standard
model \ref\baur{U.~Baur and E.~W.~N.~Glover, \np {B347} {1990} {12}.}.
Lastly, there is a two-loop diagram, in which a box of quarks (not all four
sides of which are top) produces a pair of Goldstone bosons which rescatter
through the Higgs boson into $Z_LZ_L$.  This last diagram must also be
computed to get a correct, gauge invariant answer, since it is leading in
$1/N$.

To get a rough estimate of the rate, we concentrate on the top-quark triangle,
ignoring the other two diagrams.  The amplitude for this diagram is
\eqn\tamp{
{\alpha_s ~s~\delta^{ab} \over
{2\pi\left[ v^2 - Ns\left({1\over\lambda(M)}
+\widetilde{B}(s;m_\psi,M)\right)\right]}}
\left( g^{\mu\nu} - {2 p_2^\mu p_1^\nu \over s}\right)
I(s,m^2_t)
\punct[.]}
Here $p_1$ and $p_2$ are the momenta of the two incoming gluons with
polarization vectors associated with $\mu$ and $\nu$ and colors with the $a$
and $b$ respectively, and the function $I(s, m^2_t)$ is the Feynman parameter
integral
\eqn\feyn{
I(s,m^2_t) = m^2_t \int_0^1 dx \int_0^{1-x} dy
{(1-4 x y) \over m^2_t - xys - i\epsilon}
\punct[,]}
and $m_t$ is the mass of the top quark. The contribution of this
process to $ZZ$ production is shown as the solid curve in
\nxxxii.

The irreducible background to observing the Higgs boson in $Z$ pairs
comes from the process $q \bar{q} \to ZZ$ and is shown as the dashed curves
in \nxxxii. We see that the background is more than an order of magnitude
larger than the gauge boson scattering signal and that this signal is
unobservable. Even the gluon fusion contribution to the $Z_L Z_L$
production cross section is lower than the background by a factor of four
or more, making detection of such a broad resonance problematic.

In \kny\ it was shown that the numbers of final state gauge boson pairs from
gauge boson scattering is roughly independent of $N$ if $\sqrt{N} M$
is held fixed. This is because as $N$ increases for fixed $\sqrt{N} M$, $M$
and the mass and width of the Higgs boson decrease like $1/\sqrt{N}$. The
increased production of Higgs bosons due to their smaller mass\foot{And
therefore higher gauge boson partonic luminosity \effectw.} is
approximately cancelled by the Higgs boson's smaller branching ratio into
$W$s and $Z$s. The number of signal events, therefore, is approximately
independent of $N$ and is the same as the number which would be present in
the model with $n=0$ described in ref. \Einhorn. Since the signal for gauge
boson scattering in that model is (marginally) observable \equiv\ and since
the number of $ZZ$ events is roughly independent of $n$, the authors of
ref. \kny\ argue that the signal may be observable for any $n$.

We have calculated the signal and background for the parameters chosen in
\kny, $n=8$ and $M=2500$ GeV (with a Higgs mass of approximately 485 GeV).
The results are plotted in \nviii\ and the results for $n=0$, $M=4300$ GeV
are plotted in \nz. It is true that the total number of gauge boson
scattering events is comparable in \figs{\nviii,\nz}, and even in \nxxxii.
However, the {\it background} is much greater when $n=8$ or 32 since the
corresponding Higgs is much lighter. The gauge boson scattering signal with
$n=8$ or 32 is ``hidden'' because the Higgs boson is both light and broad.
In both cases, the gluon fusion signal is substantially larger than the
gauge boson scattering signal. The signal, however, is still significantly
below the background, making detection of a broad resonance difficult, at
best.

By way of comparison, the signal for a 485 GeV standard model Higgs is
shown in \higgs. In this case, because the Higgs is relatively narrow, {\it
on the peak} the gauge boson scattering Higgs signal is comparable to the
background and the gluon fusion signal is well above the background.

There are two technical shortcomings in the calculation of the cross
sections presented above. Firstly, in computing the gauge boson scattering
signal, we have used both the equivalence theorem \equiv\ and the
effective-$W$ approximation \effectw. Strictly speaking, both of these
approximations hold only at energies above a few times the $W$ mass. Even
at 200 GeV, however, the corrections should be of order one
\effectw\ref\gunion{ J.~F.~Gunion, J.~Kalinowski, and A. Tofighi-Niaki,
\prl{57}{1986}{2351}}, whereas the background at these energies when $n =
32$ is more than an {\it order of magnitude} larger. While we cannot precisely
determine the number of gauge boson scattering events, it is clear that they
are swamped by the background. Second, in computing the gluon fusion signal,
we have only computed the contribution from a top quark triangle diagram and
have not included the contributions from the other two of the three leading
diagrams, as discussed above.  These will interfere with the contribution we
have computed.  However, we do not expect this to change any of our
conclusions.

In general, the two-gauge-boson scattering signal of the symmetry breaking
sector is not visible above the background unless the gauge boson elastic
scattering amplitudes are big. This can happen either if the symmetry
breaking sector is strongly coupled or at the peak of a narrow resonance.
We can see this just by counting coupling constants: the background ($qq
\rightarrow ZZ$) is order $g^2$ and is a two body final state, while the
signal ($qq \rightarrow qqZZ$) is naively of order $g^4$ and is a four body
final state. When the symmetry breaking sector is strongly interacting and
the final state gauge bosons are longitudinal, this naive $g^4$ gets
replaced by $g^2 a^{ij;kl}$, and the signal may compete with the
background. Since $a^{ij;kl}$ in the $O(4+32)$ model is {\it never} large,
the signal rate {\it never approaches} the background rate.

Moreover, while we have concentrated on the signal for the $ZZ$ final state,
the arguments given here should apply equally to all other two-gauge-boson
signals as well. In the $O(4+32)$ model it is likely that {\it none} of the
two-gauge-boson signals of the symmetry breaking sector may be observed over
the background.  In this model even observing all two-gauge-boson modes will
not be sufficient to detect the dynamics of electroweak symmetry breaking --
one will need to observe the pseudo-Goldstone bosons, and identify them with
symmetry breaking.

In conclusion, we see that in models of electroweak symmetry breaking with a
large number of pseudo-Goldstone bosons the longitudinal gauge boson
scattering amplitudes may be small and structureless at all energies.  In this
case we cannot necessarily rely on gauge boson pairs as a signal of the
dynamics of symmetry breaking.

\bigskip\medskip

We would like to thank Kenneth Lane for useful conversations, Elizabeth
Simmons for reading the manuscript, and Gordon Kane, Steven Naculich, and
C.-P. Yuan for sending us ref. \kny\ prior to publication. R.S.C.
acknowledges the support of an Alfred P. Sloan Foundation Fellowship, an
NSF Presidential Young Investigator Award and DOE Outstanding Junior
Investigator Award. This work was supported in part under NSF contract
PHY-9057173 and DOE contracts DE-AC02-89ER40509 and DE-FG02-91ER40676, and
by funds from the Texas National Research Laboratory Commission under grant
RGFY91B6.

\listfigs
\listrefs

\bye